\documentclass[a4paper]{article}

\usepackage{INTERSPEECH2021}
\usepackage{multirow}
\usepackage{subfigure}

\title{Improving Streaming Transformer Based ASR Under a Framework of Self-supervised Learning}
\name{Songjun Cao, Yueteng Kang, Yanzhe Fu, Xiaoshuo Xu, Sining Sun, Yike Zhang, Long Ma }
\address{
  Tencent Cloud Xiaowei, Beijing, China}
\email{\{songjuncao,yuetengkang,yanzhefu,xiaoshuoxu,siningsun,yikezhang,malonema\}@tencent.com}

\begin{document}

\maketitle
\begin{abstract}
Recently self-supervised learning has emerged as an effective approach to improve the performance of automatic speech recognition (ASR).
Under such a framework, the neural network is usually pre-trained with massive unlabeled data and then fine-tuned with limited labeled data.
However, the non-streaming architecture like bidirectional transformer is usually adopted by the neural network to achieve competitive results, which can not be used in streaming scenarios.
In this paper, we mainly focus on improving the performance of streaming transformer under the self-supervised learning framework.
Specifically,
we propose a novel two-stage training method during fine-tuning, which combines knowledge distilling and self-training.
The proposed training method achieves 16.3\% relative word error rate (WER) reduction on Librispeech noisy test set.
Finally, by only using the 100h clean subset of Librispeech as the labeled data and the rest (860h) as the unlabeled data, our streaming transformer based model obtains competitive WERs 3.5/8.7 on Librispeech clean/noisy test sets.

\end{abstract}
\noindent\textbf{Index Terms}: speech recognition, streaming transformer, self-supervised learning, knowledge distilling

\section{Introduction}
Self-supervised learning achieves great success in natural language processing \cite{peters2018deep,radford2018improving,devlin2018bert}, automatic speech recognition \cite{Schneider2019,DBLP:conf/iclr/BaevskiSA20,9054224,baevski2020wav2vec,zhang2020pushing} and computer vision \cite{henaff2020data,chen2020simple,misra2020self} tasks.
For speech recognition, self-supervised learning enables the model to learn general data representations from unlabeled examples. Fine-tuned by limited labeled data, the model can achieve good performance. 
It is very promising and valuable for many application scenarios, as labeled data is much harder to get than unlabeled data.

To achieve better results, the model used by self-supervised learning usually adopts bidirectional transformer architecture with self attention \cite{vaswani2017attention,devlin2018bert}.
It can not be used for streaming scenarios, as the model needs to attend the full sequence.
In order to solve this problem, there are some streaming solutions proposed.
\cite{povey2018time} realizes a streaming self attention layer for ASR by limiting number of frames to the left and right.
As the reception field grows linearly with the number of transformer layers, there will be a large latency.
\cite{chen2020developing} splits the input into small chunks with a specified chunk size.
By the special attention mask design, it allows the left context linearly increase while forbidding the right reception field to grow, which is easy to control the latency.
A block processing method is proposed by \cite{dong2019self} which segments the entire utterance into several chunks.
Each of the chunks has three parts: the past part, the current part, and the future part.
Compared to \cite{chen2020developing}, the future part provides more future frames given a fixed chunk size at the cost of introducing more computation.
Based on \cite{dong2019self}, a streaming transformer with augmented memory is proposed in \cite{Wu2020} to reduce latency and the self-attention computation.
Emformer is introduced by \cite{shi2020emformer} to further improve the computation efficiency.

Under the framework of self-supervised learning, we can improve our model's performance furthermore by utilizing self-training \cite{kahn2020self}.
However, previous works \cite{zhang2020pushing,xu2020self} introduce self-training to bidirectional transformer based models.
It may be a good choice for streaming transformer based models too.
Besides the self-training method, knowledge distillation \cite{romero2014fitnets} is another choice to utilize the unlabeled data as it uses the teacher model's output as the soft label. However, it is usually used to transfer the knowledge of a large model to a small model \cite{jiao2019tinybert}.

In this paper, we construct a streaming speech recognition system based on \cite{baevski2020wav2vec}, which is one of the most effective frameworks for self-supervised learning.
We try to introduce streaming transformer during fine-tuning and investigate three types of streaming transformer in our experiments, which are described in Section~\ref{section:streaming}.
Under the average latency of 480 ms constraint, Block Transformer achieves the best result.
To the best of our knowledge, this is the first work to explore streaming transformer under a self-supervised learning framework.

Next, we make some efforts to improve the performance of streaming transformer based model.
Here we are focusing on the semi-supervised method which makes use of unlabeled data.
A two-stage training method is proposed in this paper.

\section{Self-supervised learning} \label{section:self}

In this paper, we experiment with the recently introduced wav2vec 2.0 model \cite{baevski2020wav2vec}.
This model is comprised of three parts.
The first part is a multi-layer convolutional feature encoder $f: \mathcal{X} \rightarrow \mathcal{Z}$, which maps the raw audio $\mathcal{X}$ to latent speech representations $\mathbf{z}_1,...,\mathbf{z}_T$. 
Each of $\mathbf{z}_t$ represents about 25ms of audio strode by 20ms.
Then the $\mathcal{Z}$ will be fed to the second part named  Transformer $g:\mathcal{Z}\rightarrow \mathcal{C}$ whose output stands for context representation $\mathbf{c}_1,...,\mathbf{c}_T$.
The third part is quantization module $\mathcal{Z}\rightarrow \mathcal{Q}$ which discretizes the output of the feature encoder and gets $\mathbf{q}_1,...,\mathbf{q}_T$, which represent the targets during training.

The model is trained by solving a contrastive task $\mathcal{L}_m$ which requires identifying the true quantized latent speech representation $\mathbf{q}_t$ for a masked time step within a set of distractors $\mathbf{Q_t}$.
The loss is defined as:
\begin{equation}
\mathcal{L}_m=-log\frac{\exp(sim(\mathbf{c}_t, \mathbf{q}_t))}{\sum_{\tilde{\mathbf{q}}\sim \mathbf{Q}_t} \exp(sim(\mathbf{c_t}, \tilde{\mathbf{q}})}
\end{equation}
where $sim(\mathbf{a}, \mathbf{b})$ denotes cosine similarity.
Besides, the loss is augmented by a codebook diversity loss and L2 penalty over the outputs of the feature encoder.

Finally, the pre-trained models are fine-tuned for speech recognition by adding a randomly initialized linear projection layer on top of the transformer network. Models are optimized by minimizing a Connectionist Temporal Classification(CTC) loss \cite{graves2006connectionist}.

\section{Streaming transformer} \label{section:streaming}

The transformer architecture used in wav2vec 2.0 model follows \cite{devlin2018bert}, which adopts the attention mechanism to capture the sequence information. 
The input vector $\mathbf{x}_t$ is projected into three parts named query $\mathbf{q}_t$, key $\mathbf{k}_t$ and value $\mathbf{v}_t$.
The attention part can be written as,
\begin{equation}
	\begin{split}
	\alpha_{t,\tau}&=\text{Softmax}(\beta \mathbf{q}^T_t \mathbf{k}_{\tau}) \\
	&=\frac{\exp(\beta(\mathbf{W}_q\mathbf{x}_t)^T(\mathbf{W}_k\mathbf{x}_{\tau}))}{\sum_{\tau'}\exp(\beta(\mathbf{W}_q\mathbf{x}_t)^T(\mathbf{W}_k\mathbf{x}_{\tau'}))} \\
	\mathbf{z}_t &= \sum_{\tau} \alpha_{t\tau} \mathbf{v}_{\tau} \\
	&= \sum_{\tau} \alpha_{t\tau}\mathbf{W}_v \mathbf{x}_{\tau}
	\end{split}
\end{equation}
where $\beta=\frac{1}{\sqrt{d}}$ is a scaling factor.

Then multi-head attention(MHA) will be applied to further improve the model capacity. 
The MHA can be calculated only when the entire inputs are ready, which can not be used in the streaming speech recognition scenarios.
However, we can apply a special attention mask on the attention weight matrix ${\alpha_{t,\tau}}$ to determine the range of input sequence involved for computation.
In this way, we can get streaming transformer architecture.

\begin{figure}[htbp]
    \centering
    \includegraphics[width=\linewidth]{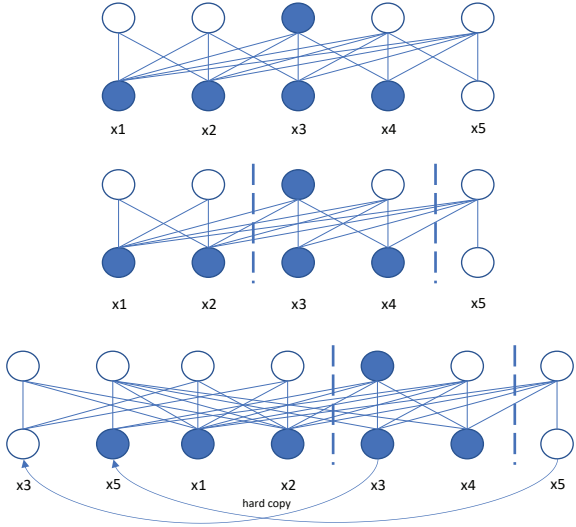}
    \caption{ Illustration of the reception field of position x3 for three streaming transformers whose left context is not limited. The top one indicates Time-restricted Transformer whose right context is 1 frame. The middle one stands for Chunk Transformer whose chunk size is 2 frames. The bottom one indicates Block Transformer whose chunk size is 2 frames and future size is 1 frame.
    }
    \label{fig:streaming}
\end{figure}

Figure \ref{fig:streaming} gives the reception field of three streaming transformers. 
Because the left context does not influence the latency, we only care about the right context here. 
The Time-restricted Transformer is based on \cite{povey2018time}, whose upper layers can get more right context as its reception field grows linearly with the number of layers.
The Chunk Transformer is defined in \cite{chen2020developing}, whose reception field varies among the time dimension.
In one chunk, the left side can see more future frames than the right side.
Block Transformer is similar to Chunk Transformer except introducing future part from \cite{dong2019self} and hard copy from \cite{shi2020emformer}.
For the Block Transformer, every frame of a chunk can see at least $F$ future frames, where $F$ stands for the size of the future part.

\section{Two-stage training method}

\begin{figure}[htbp]
    \centering
    \includegraphics[width=\linewidth]{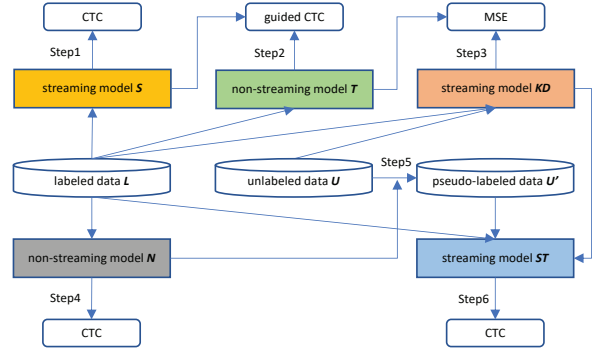}
    \caption{ Illustration of our two-stage training method.
    }
    \label{fig:two-stage}
\end{figure}

Instead of using CTC loss to do the fine-tuning on labeled data directly, we propose a two-stage training method to better utilize the unlabeled data, which is shown in Figure \ref{fig:two-stage}.
Suppose we have a pre-trained model named $P$ which is non-streaming model, labeled data set $L$, unlabeled data set $U$ and a language model (LM), our method can be described as the following procedure:
\begin{itemize}
\item Step1: apply the attention mask to pre-trained model $P$ and fine-tune it on $L$ with CTC loss to get the streaming model $S$.
\item Step2: fine-tune pre-trained model $P$ on $L$ with $S$ and guided CTC loss to get the non-streaming model $T$.
\item Step3: apply the attention mask to pre-trained model $P$ and fine-tune it on $L$ and $U$ with distillation loss and $T$ to get the streaming model $KD$.
\item Step4: fine-tune pre-trained model $P$ on $L$ with CTC loss to get the non-streaming model $N$.
\item Step5: decode $U$ with $N$ and LM to get the pseudo-labeled data set $U'$.
\item Step6: fine-tune model $KD$ on $L$ and $U'$ with CTC loss to get the final streaming model $ST$.
\end{itemize}

\subsection{Knowledge distillation}
Knowledge distillation aims to transfer the knowledge of a large teacher model to a small student model.
The student model is trained to mimic the behaviors of the teacher model.
Motivated by \cite{jiao2019tinybert}, we treat the non-streaming model as the teacher model and the streaming model as the student model.

As discussed in \cite{kurata2019guiding,kurata2020knowledge}, CTC models suffer from the disagreement of spike timings during knowledge distillation from non-streaming model to streaming model.
So we use $T$ as our teacher model which is trained with guided CTC loss \cite{kurata2019guiding} as follows:
\begin{equation}
\mathcal{L}=\mathcal{L}_{CTC}+\alpha \mathcal{L}_G \\
\label{eq:guide}
\end{equation}
\begin{equation}
\mathcal{L}_G=-1\cdot \sum M(\mathbf{X})\circ P(\mathbf{X})
\end{equation}

where $\circ$ denotes element-wise product, $\alpha$ is the hyperparameter, $P(\mathbf{X})$ denotes posteriors obtained by non-streaming model $T$, $M(\mathbf{X})$ is a mask which can be obtained from posteriors of streaming model $S$ by setting a 1 at the output symbol with the highest posterior and 0 at other symbols at each time index. In cases where the blank symbol has the highest posterior, we set 0 for all symbols at this time index.

The distillation loss can be defined as follows:
\begin{equation}
\mathcal{L}_{distill}=\sum_{i}MSE(\mathbf{H}^S_i, \mathbf{H}^T_i)
\end{equation}
where $i$ denotes the $i$-th transformer layer, $\mathbf{H}_i$ denotes the $i$-th transformer layer's output from the teacher model $T$ or the student model $KD$.
Finally we can get the streaming model $KD$.

\subsection{Self-training}
At the second stage, we adopt the pseudo-labeling strategy of \cite{synnaeve2019end,kahn2020self}.
As \cite{doutre2020improving}, we first fine-tune an initial non-streaming model $N$ on the available labeled data.
Then we use $N$ and LM to label the unlabeled data and get the pseudo-labeled data.
Finally, the model $KD$ from the first stage will be fine-tuned on the original labeled data and pseudo-labeled data.
A simple and effective self-training method as \cite{xu2020self} is adopted here. We opt for a single iteration that is computationally less while still enabling the fairness of comparison.
CTC loss is used during training.

\section{Experiment}
\subsection{Experiment setup}
In this paper, speech audio of the Librispeech corpus (LS-960) \cite{panayotov2015librispeech} is used as our training data.
During the pre-training, we use all the data without transcriptions as the training set.
During the fine-tuning, we use the train-clean-100 subset comprising 100h as labeled data and the left 860 hours as unlabeled data.
We evaluate our models on Librispeech dev-other/clean and test-clean/other sets.

For the model structure, our experiments are based on the BASE model of \cite{baevski2020wav2vec}, which contains 12 transformer layers. It is pre-trained with 960 hours Librispeech audio with the same setup as \cite{baevski2020wav2vec}. During fine-tuning, we optimize with Adam and a tri-stage rate schedule where the learning rate is warmed up for the first 10\% of updates, held constant for the next 40\% and then linearly decayed for the remainder. The peak of learning rate is set to 2e-5 and the max number of update is set to 80000.
Our dictionary contains 29 tokens, including 26 English characters and 3 special symbols.

For decoding, the standard Librispeech 4-gram language model is used to do beam-search for all the results. The beam size is fixed to 500. Models trained by self-training use 1.74 and -0.52 as LM weight and word insertion penalty, while others use 2.15 and -0.52 as suggested by \cite{baevski2020wav2vec}. 

\subsection{Baseline}
\begin{table}[htp]
  \caption{WER results of our baseline model. \textbf{GN} and \textbf{BN} indicate group normalization and batch normalization. \textbf{SC} and \textbf{CC} indicate symmetric convolution and causal convolution.}
  \label{tab:baseline}
  \centering
  \begin{tabular}{ccccccc}
    \toprule
    \multirow{2}{*}{\textbf{Model}} & \multirow{2}{*}{\textbf{Norm}} & \multirow{2}{*}{\textbf{Conv}} & \multicolumn{2}{c}{\textbf{dev}} & \multicolumn{2}{c}{\textbf{test}} \\
    & & & \textbf{clean} & \textbf{other} & \textbf{clean} & \textbf{other} \\
    \midrule
    \cite{baevski2020wav2vec} & GN & SC & 2.7 & 7.9 & 3.4 & 8.0  \\
    N1 & GN & SC & 2.9 & 8.1 & 3.3 & 8.1  \\
    N2 & BN & SC & 2.9 & 8.1 & 3.3 & 8.4  \\
    N3 & BN & CC & 2.9 & 8.5 & 3.5 & 8.4  \\
    \bottomrule
  \end{tabular}
\end{table}

First, we try to reproduce the result of \cite{baevski2020wav2vec}, which is $N1$ of Table \ref{tab:baseline}.
As the original wav2vec 2.0 model contains group normalization in the feature encoder part and a convolutional layer with kernel size 128 \cite{mohamed2019transformers} before the transformer, which are not suitable in streaming system. 
Here we use batch normalization \cite{ioffe2015batch} and a causal convolutional layer with kernel size 24 to replace them.
The result of the modified model is displayed in the last row of Table \ref{tab:baseline}, which is a little worse than the original model.
\subsection{Results of streaming transformer}

\begin{table}[htp]
  \caption{WER results of streaming transformer based models whose EIL is fixed at 480 ms. \textbf{N3} indicates bidirectional transformer, \textbf{S1} indicates Time-restricted Transformer, \textbf{S2} indicates Chunk Transformer, \textbf{S3} and \textbf{S4} indicate Block Transformer. \textbf{C} and \textbf{F} are the chunk size and future size (in millisecond).}
  \label{tab:streaming}
  \centering
  \begin{tabular}{ccccccc}
    \toprule
    \multirow{2}{*}{\textbf{Model}} & \multirow{2}{*}{\textbf{C}} & \multirow{2}{*}{\textbf{F}} & \multicolumn{2}{c}{\textbf{dev}} & \multicolumn{2}{c}{\textbf{test}} \\
    & & & \textbf{clean} & \textbf{other} & \textbf{clean} & \textbf{other} \\
    \midrule
    N3 & - & - & 2.9 & 8.5 & 3.5 & 8.4  \\
    S1 & - & - & 3.9 & 13.0 & 4.4 & 13.0  \\
    S2 & 960 & 0 & 3.5 & 11.4 & 3.9 & 11.4  \\
    S3 & 480 & 240 & 3.4 & 10.5 & 3.9 & 10.6  \\
    S4 & 240 & 360 & 3.5 & 10.3 & 3.9 & 10.4  \\
    \bottomrule
  \end{tabular}
\end{table}

To measure the latency introduced by streaming transformer, we use algorithmic latency induced by the encoder (EIL) proposed in \cite{shi2020emformer}.
Here we focus on the accuracy of streaming transformer, so the limited left context and computation optimization will be explored in future work.

First, we try to introduce streaming transformer during fine-tuning, while bidirectional transformer is still used during pre-training.
For comparison, we fix the EIL of all the transformers at 480 ms.
$S1$ stands for Time-restricted Transformer whose right context of every transformer layer is set to 2 frames.
$S2$ indicates Chunk Transformer whose chunk size is set to 48 frames.
$S3$ and $S4$ denote Block Transformer, whose chunk size and future size are set to 24/12 frames and 12/18 frames.
As we can see from Table \ref{tab:streaming}, the Block Transformer achieves the best result.
We argue that the future part introduced by Block Transformer is crucial for streaming transformer as it can guarantee $F$ frames' right context even for the rightest frame in a chunk.
Compared with the bidirectional transformer, our streaming transformer's WER is increased from 8.4 to 10.4 on the test-other set.

Next, we try to introduce streaming transformer during both pre-training and fine-tuning, which may be good from the perspective of matching.
But the results show that it doesn't work well, as we find the validation loss of streaming transformer becomes worse during pre-training.
We will leave it for future exploration.

Our experiments below are based on streaming transformer $S4$.


\subsection{Results of two-stage training method}

\begin{table}[htp]
  \caption{WER results of knowledge distillation and self-training.
  \textbf{KD} indicates knowledge distillation and \textbf{ST} indicates self-training.}
  \label{tab:semi}
  \centering
  \begin{tabular}{ccccccc}
    \toprule
    \multirow{2}{*}{\textbf{Model}} & \multirow{2}{*}{\textbf{KD}} & \multirow{2}{*}{\textbf{ST}} & \multicolumn{2}{c}{\textbf{dev}} & \multicolumn{2}{c}{\textbf{test}} \\
    & & & \textbf{clean} & \textbf{other} & \textbf{clean} & \textbf{other} \\
    \midrule
    N1 & - & - & 2.9 & 8.1 & 3.3 & 8.1  \\
    N4 & N & Y & 2.9 & 7.4 & 3.2 & 7.6  \\
    S4 & - & - & 3.5 & 10.3 & 3.9 & 10.4  \\
    S5 & Y & N & 3.3 & 9.6 & 3.7 & 9.8  \\
    S6 & N & Y & 3.3 & 8.6 & 3.6 & 8.9  \\
    S7 & Y & Y & 3.2 & 8.5 & 3.5 & 8.7  \\
    \bottomrule
  \end{tabular}
\end{table}

We perform the knowledge distillation on 960 hours of data without labels for 40000 updates.
We do the distillation on three transformer layers which are Layer4 Layer8 and Layer12.
As for the linear projection layer on top of the transformer, we find that directly using that of $S4$ works well compared to extra training.

For the self-training, we use $N1$ of Table \ref{tab:baseline} and the standard Librispeech 4-gram language model to decode the unlabeled 860 hours data.
The total training data is composed of 100 hours of labeled data and 860 hours of pseudo-labeled data.

Results are shown in Table \ref{tab:semi}.
For a fair comparison, self-training in $N4$ and $S6$ is trained for 120000 updates while that in $S7$ is trained for 80000 updates.
Compared with the baseline model named $S4$, our two-stage training method ($S7$) gets 16.3\% relative WER reduction, which is better than knowledge distillation ($S5$) and self-training ($S6$).
For self-training, our streaming model is supervised by the pseudo labels generated by beam-search.
For knowledge distillation, it is supervised by outputs of the non-streaming model's intermediate transformer layers, which are soft labels.
As the two kinds of supervised information are different, the two methods are complementary to some extent.


\begin{figure}
\centering
\subfigure[Streaming transformer]{\label{fig:spike1}\includegraphics[width=\linewidth]{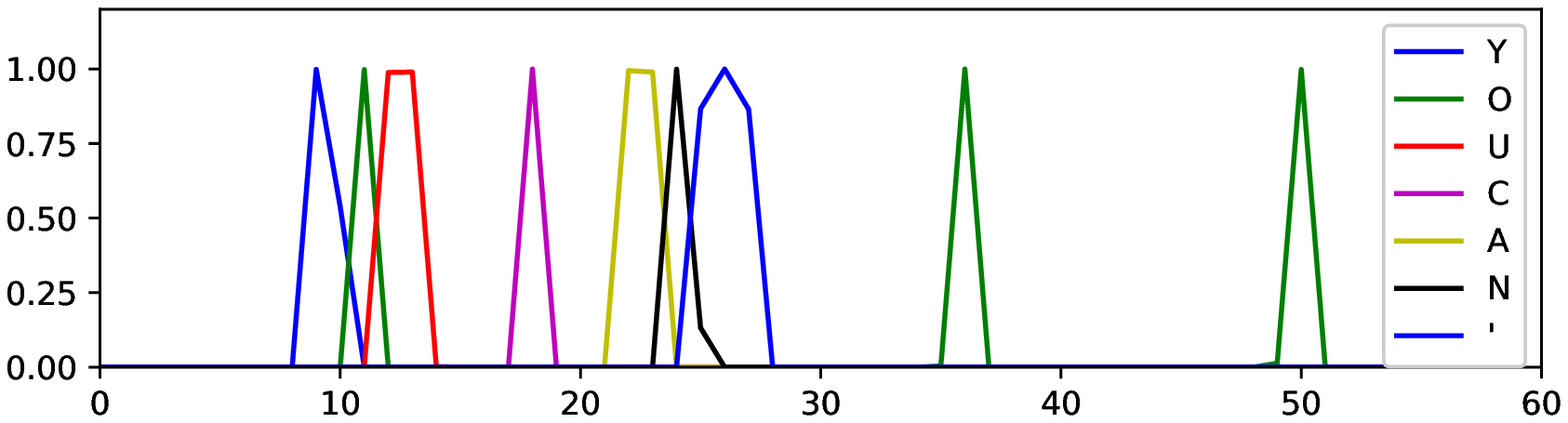}} 
\\ 
\centering
\subfigure[Bidirectional transformer]{\label{fig:spike2}\includegraphics[width=\linewidth]{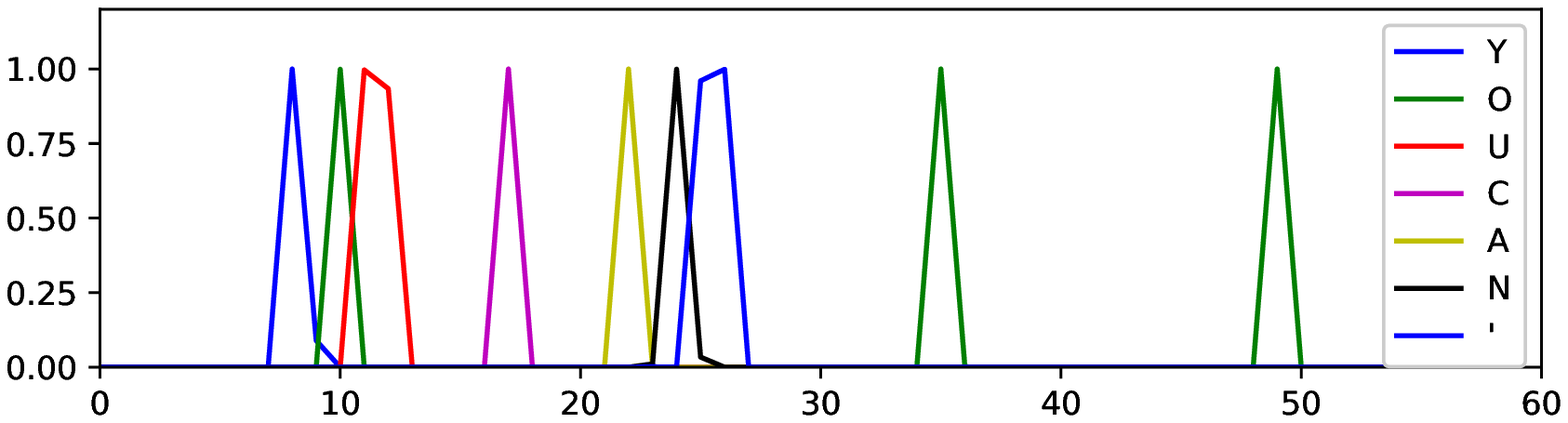}}
\\
\centering
\subfigure[Bidirectional transformer guided by streaming transformer]{\label{fig:spike3}\includegraphics[width=\linewidth]{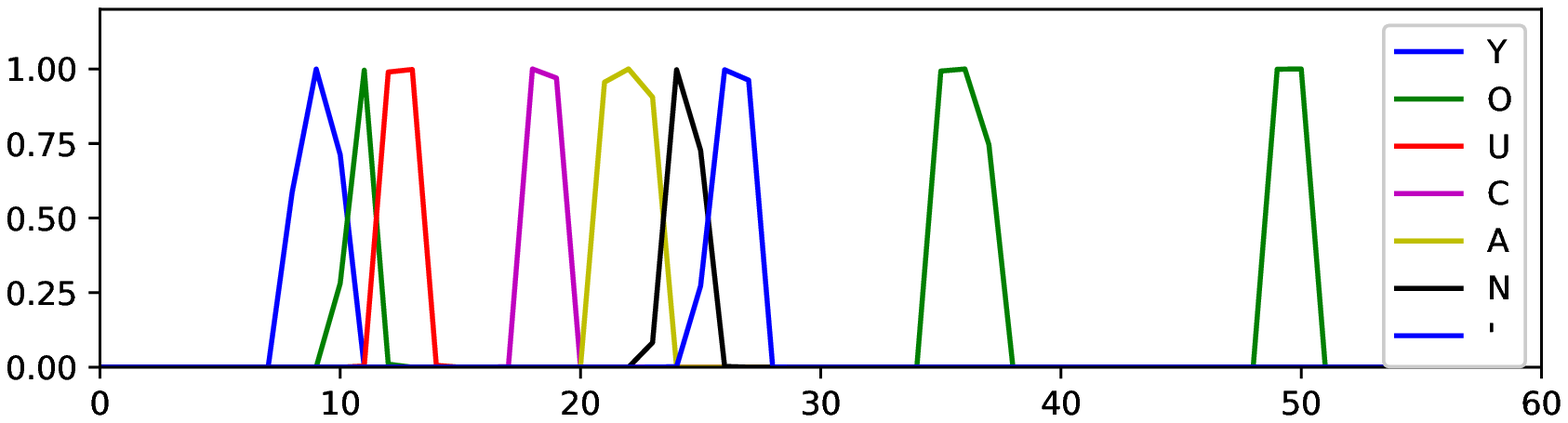}}
\caption{Illustration of spikes of CTC's output. The horizontal axis represents speech frames (the frame shift is 20 ms). The vertical axis represents posteriors.}
\label{fig:spike}
\end{figure}

For further discussion, we try to use self-training on the non-streaming model too, whose result is $N4$ in Table \ref{tab:semi}. Knowledge distillation is not conducted for the non-streaming model as the teacher model and student model are the same.
After training with more unlabeled data, the gap between non-streaming model and streaming model can be reduced from 22.1\% to 12.6\%.
One explanation is that the streaming model can benefit from both knowledge distillation and self-training while the non-streaming model mainly benefits from self-training.

\subsection{Ablation of knowledge distillation}

For knowledge distillation, we investigate 4 teacher models, as depicted in Table \ref{tab:teacher}.
One is trained by CTC loss, while the rest is trained by guided CTC loss.
The teacher model achieves better results with smaller $\alpha$, which is similar to \cite{kurata2020knowledge}.

\begin{table}[htp]
  \caption{WER results of teacher models used by distillation. $\alpha$ is the hyperparameter in Eq. \ref{eq:guide}.}
  \label{tab:teacher}
  \centering
  \begin{tabular}{ccccccc}
    \toprule
    \multirow{2}{*}{\textbf{Model}} & \multirow{2}{*}{\textbf{$\alpha$}} & \multicolumn{2}{c}{\textbf{dev}} & \multicolumn{2}{c}{\textbf{test}} \\
    & & \textbf{clean} & \textbf{other} & \textbf{clean} & \textbf{other} \\
    \midrule
    T1 & 0 & 2.9 & 8.5 & 3.5 & 8.4  \\
    T2 & 1 & 3.3 & 8.8 & 3.8 & 8.8  \\
    T3 & 0.1 & 3.1 & 8.6 & 3.7 & 8.6  \\
    T4 & 0.01 & 3.1 & 8.4 & 3.6 & 8.4  \\
    \bottomrule
  \end{tabular}
\end{table}

\begin{table}[htp]
  \caption{WER results of distillation models.}
  \label{tab:distill}
  \centering
  \begin{tabular}{cccccc}
    \toprule
    \multirow{2}{*}{\textbf{Model}} & \multirow{2}{*}{\textbf{Teacher}} & \multicolumn{2}{c}{\textbf{dev}} & \multicolumn{2}{c}{\textbf{test}} \\
    & & \textbf{clean} & \textbf{other} & \textbf{clean} & \textbf{other} \\
    \midrule
    KD1 & T1 & 3.3 & 9.9 & 3.7 & 10.0  \\
    KD2 & T2 & 3.5 & 9.7 & 4.0 & 10.0  \\
    KD3 & T3 & 3.4 & 9.8 & 3.7 & 9.9  \\
    KD4 & T4 & 3.3 & 9.6 & 3.7 & 9.8  \\
    \bottomrule
  \end{tabular}
\end{table}

Table \ref{tab:distill} shows the results for the distillation.
$KD4$ achieves the best result, which corresponds to $S5$ in Table \ref{tab:semi}.
We can conclude that our distillation benefits from the guided CTC loss.
However, the gain of guided CTC loss is not as much as \cite{kurata2019guiding}, which can be explained by Figure \ref{fig:spike}.
As our streaming transformer can see some future frames, the mismatch of spiking times between streaming transformer and bidirectional transformer is not as obvious as that in \cite{kurata2019guiding}. 

For greater clarity, models $S4$/$T4$/$S5$/$N1$/$S7$ in our experiments correspond to models $S$/$T$/$KD$/$N$/$ST$ in Figure \ref{fig:two-stage}.


\section{Conclusion}
In this paper, we explore the performance of streaming speech recognition system which is based on the wav2vec 2.0 framework.
First, we find that introducing streaming transformer during the fine-tuning works well.
Then, we investigate the performance of different streaming transformers and find that Block Transformer is the best choice.
Finally, we try to improve the performance of the streaming transformer based model by knowledge distillation and self-training during fine-tuning.

However, there are still some issues worth exploring in the future. One of them is to introduce the streaming transformer during the pre-training. Designing more efficient structure for the streaming transformer is another interesting topic.

\bibliographystyle{IEEEtran}

\bibliography{mybib}

\begin{thebibliography}{10}
\providecommand{\url}[1]{#1}
\csname url@samestyle\endcsname
\providecommand{\newblock}{\relax}
\providecommand{\bibinfo}[2]{#2}
\providecommand{\BIBentrySTDinterwordspacing}{\spaceskip=0pt\relax}
\providecommand{\BIBentryALTinterwordstretchfactor}{4}
\providecommand{\BIBentryALTinterwordspacing}{\spaceskip=\fontdimen2\font plus
\BIBentryALTinterwordstretchfactor\fontdimen3\font minus
  \fontdimen4\font\relax}
\providecommand{\BIBforeignlanguage}[2]{{%
\expandafter\ifx\csname l@#1\endcsname\relax
\typeout{** WARNING: IEEEtran.bst: No hyphenation pattern has been}%
\typeout{** loaded for the language `#1'. Using the pattern for}%
\typeout{** the default language instead.}%
\else
\language=\csname l@#1\endcsname
\fi
#2}}
\providecommand{\BIBdecl}{\relax}
\BIBdecl

\bibitem{peters2018deep}
M.~E. Peters, M.~Neumann, M.~Iyyer, M.~Gardner, C.~Clark, K.~Lee, and
  L.~Zettlemoyer, ``Deep contextualized word representations,'' \emph{arXiv
  preprint arXiv:1802.05365}, 2018.

\bibitem{radford2018improving}
A.~Radford, K.~Narasimhan, T.~Salimans, and I.~Sutskever, ``Improving language
  understanding by generative pre-training,'' 2018.

\bibitem{devlin2018bert}
J.~Devlin, M.-W. Chang, K.~Lee, and K.~Toutanova, ``Bert: Pre-training of deep
  bidirectional transformers for language understanding,'' \emph{arXiv preprint
  arXiv:1810.04805}, 2018.

\bibitem{Schneider2019}
S.~Schneider, A.~Baevski, R.~Collobert, and M.~Auli, ``{wav2vec: Unsupervised
  Pre-Training for Speech Recognition},'' in \emph{Proc. Interspeech 2019},
  2019, pp. 3465--3469.

\bibitem{DBLP:conf/iclr/BaevskiSA20}
A.~Baevski, S.~Schneider, and M.~Auli, ``vq-wav2vec: Self-supervised learning
  of discrete speech representations,'' in \emph{8th International Conference
  on Learning Representations, {ICLR} 2020, Addis Ababa, Ethiopia, April 26-30,
  2020}, 2020.

\bibitem{9054224}
A.~{Baevski} and A.~{Mohamed}, ``{Effectiveness of Self-Supervised Pre-Training
  for ASR},'' in \emph{ICASSP 2020 - 2020 IEEE International Conference on
  Acoustics, Speech and Signal Processing (ICASSP)}, 2020, pp. 7694--7698.

\bibitem{baevski2020wav2vec}
A.~Baevski, Y.~Zhou, A.~Mohamed, and M.~Auli, ``wav2vec 2.0: A framework for
  self-supervised learning of speech representations,'' \emph{Advances in
  Neural Information Processing Systems}, vol.~33, 2020.

\bibitem{zhang2020pushing}
Y.~Zhang, J.~Qin, D.~S. Park, W.~Han, C.-C. Chiu, R.~Pang, Q.~V. Le, and Y.~Wu,
  ``Pushing the limits of semi-supervised learning for automatic speech
  recognition,'' \emph{arXiv preprint arXiv:2010.10504}, 2020.

\bibitem{henaff2020data}
O.~Henaff, ``Data-efficient image recognition with contrastive predictive
  coding,'' in \emph{International Conference on Machine Learning}.\hskip 1em
  plus 0.5em minus 0.4em\relax PMLR, 2020, pp. 4182--4192.

\bibitem{chen2020simple}
T.~Chen, S.~Kornblith, M.~Norouzi, and G.~Hinton, ``A simple framework for
  contrastive learning of visual representations,'' in \emph{International
  conference on machine learning}.\hskip 1em plus 0.5em minus 0.4em\relax PMLR,
  2020, pp. 1597--1607.

\bibitem{misra2020self}
I.~Misra and L.~v.~d. Maaten, ``Self-supervised learning of pretext-invariant
  representations,'' in \emph{Proceedings of the IEEE/CVF Conference on
  Computer Vision and Pattern Recognition}, 2020, pp. 6707--6717.

\bibitem{vaswani2017attention}
A.~Vaswani, N.~Shazeer, N.~Parmar, J.~Uszkoreit, L.~Jones, A.~N. Gomez,
  L.~Kaiser, and I.~Polosukhin, ``Attention is all you need,'' in \emph{NIPS},
  2017.

\bibitem{povey2018time}
D.~Povey, H.~Hadian, P.~Ghahremani, K.~Li, and S.~Khudanpur, ``{A
  time-restricted self-attention layer for ASR},'' in \emph{2018 IEEE
  International Conference on Acoustics, Speech and Signal Processing
  (ICASSP)}.\hskip 1em plus 0.5em minus 0.4em\relax IEEE, 2018, pp. 5874--5878.

\bibitem{chen2020developing}
X.~Chen, Y.~Wu, Z.~Wang, S.~Liu, and J.~Li, ``Developing real-time streaming
  transformer transducer for speech recognition on large-scale dataset,''
  \emph{arXiv preprint arXiv:2010.11395}, 2020.

\bibitem{dong2019self}
L.~Dong, F.~Wang, and B.~Xu, ``{Self-attention aligner: A latency-control
  end-to-end model for ASR using self-attention network and chunk-hopping},''
  in \emph{ICASSP 2019-2019 IEEE International Conference on Acoustics, Speech
  and Signal Processing (ICASSP)}.\hskip 1em plus 0.5em minus 0.4em\relax IEEE,
  2019, pp. 5656--5660.

\bibitem{Wu2020}
\BIBentryALTinterwordspacing
C.~Wu, Y.~Wang, Y.~Shi, C.-F. Yeh, and F.~Zhang, ``{Streaming Transformer-Based
  Acoustic Models Using Self-Attention with Augmented Memory},'' in \emph{Proc.
  Interspeech 2020}, 2020, pp. 2132--2136. [Online]. Available:
  \url{http://dx.doi.org/10.21437/Interspeech.2020-2079}
\BIBentrySTDinterwordspacing

\bibitem{shi2020emformer}
Y.~Shi, Y.~Wang, C.~Wu, C.-F. Yeh, J.~Chan, F.~Zhang, D.~Le, and M.~Seltzer,
  ``Emformer: Efficient memory transformer based acoustic model for low latency
  streaming speech recognition,'' \emph{arXiv preprint arXiv:2010.10759}, 2020.

\bibitem{kahn2020self}
J.~Kahn, A.~Lee, and A.~Hannun, ``Self-training for end-to-end speech
  recognition,'' in \emph{ICASSP 2020-2020 IEEE International Conference on
  Acoustics, Speech and Signal Processing (ICASSP)}.\hskip 1em plus 0.5em minus
  0.4em\relax IEEE, 2020, pp. 7084--7088.

\bibitem{xu2020self}
Q.~Xu, A.~Baevski, T.~Likhomanenko, P.~Tomasello, A.~Conneau, R.~Collobert,
  G.~Synnaeve, and M.~Auli, ``Self-training and pre-training are complementary
  for speech recognition,'' \emph{arXiv preprint arXiv:2010.11430}, 2020.

\bibitem{romero2014fitnets}
A.~Romero, N.~Ballas, S.~E. Kahou, A.~Chassang, C.~Gatta, and Y.~Bengio,
  ``Fitnets: Hints for thin deep nets,'' \emph{arXiv preprint arXiv:1412.6550},
  2014.

\bibitem{jiao2019tinybert}
X.~Jiao, Y.~Yin, L.~Shang, X.~Jiang, X.~Chen, L.~Li, F.~Wang, and Q.~Liu,
  ``Tinybert: Distilling bert for natural language understanding,'' \emph{arXiv
  preprint arXiv:1909.10351}, 2019.

\bibitem{graves2006connectionist}
A.~Graves, S.~Fern{\'a}ndez, F.~Gomez, and J.~Schmidhuber, ``Connectionist
  temporal classification: labelling unsegmented sequence data with recurrent
  neural networks,'' in \emph{Proceedings of the 23rd international conference
  on Machine learning}, 2006, pp. 369--376.

\bibitem{kurata2019guiding}
G.~Kurata and K.~Audhkhasi, ``Guiding ctc posterior spike timings for improved
  posterior fusion and knowledge distillation,'' \emph{Proc. Interspeech 2019},
  pp. 1616--1620, 2019.

\bibitem{kurata2020knowledge}
G.~Kurata and G.~Saon, ``Knowledge distillation from offline to streaming rnn
  transducer for end-to-end speech recognition,'' \emph{Proc. Interspeech
  2020}, pp. 2117--2121, 2020.

\bibitem{synnaeve2019end}
G.~Synnaeve, Q.~Xu, J.~Kahn, T.~Likhomanenko, E.~Grave, V.~Pratap, A.~Sriram,
  V.~Liptchinsky, and R.~Collobert, ``{End-to-end ASR: from supervised to
  semi-supervised learning with modern architectures},'' \emph{arXiv preprint
  arXiv:1911.08460}, 2019.

\bibitem{doutre2020improving}
T.~Doutre, W.~Han, M.~Ma, Z.~Lu, C.-C. Chiu, R.~Pang, A.~Narayanan, A.~Misra,
  Y.~Zhang, and L.~Cao, ``Improving streaming automatic speech recognition with
  non-streaming model distillation on unsupervised data,'' \emph{arXiv preprint
  arXiv:2010.12096}, 2020.

\bibitem{panayotov2015librispeech}
V.~Panayotov, G.~Chen, D.~Povey, and S.~Khudanpur, ``{Librispeech: an ASR
  corpus based on public domain audio books},'' in \emph{2015 IEEE
  international conference on acoustics, speech and signal processing
  (ICASSP)}.\hskip 1em plus 0.5em minus 0.4em\relax IEEE, 2015, pp. 5206--5210.

\bibitem{mohamed2019transformers}
A.~Mohamed, D.~Okhonko, and L.~Zettlemoyer, ``{Transformers with convolutional
  context for ASR},'' \emph{arXiv preprint arXiv:1904.11660}, 2019.

\bibitem{ioffe2015batch}
S.~Ioffe and C.~Szegedy, ``Batch normalization: Accelerating deep network
  training by reducing internal covariate shift,'' in \emph{International
  conference on machine learning}.\hskip 1em plus 0.5em minus 0.4em\relax PMLR,
  2015, pp. 448--456.

\end{thebibliography}

\end{document}